\documentclass[a4paper]{jpconf}
\usepackage{graphicx}

\usepackage{amssymb}

\begin{document}

%\title{Sensitivity to neutron captures and $\beta$-decays
%of the s-process in massive stars at low metallicities}
\title{Sensitivity to neutron captures and $\beta$-decays
of the enhanced s-process in rotating massive stars at low metallicities}

\author{N~Nishimura$^{1}$, R~Hirschi$^{1,2}$ and T~Rauscher$^{3,4}$}
\address{${}^1$Astrophysics, Faculty of Natural Sciences, Keele University, Keele ST5~5BG, UK}
\address{${}^2$WPI Kavli IPMU, The University of Tokyo, Kashiwa 277--8583, Japan}
\address{${}^3$Department of Physics, University of Basel, 4052 Basel, Switzerland}
\address{${}^4$Centre for Astrophysics Research, University of Hertfordshire, Hatfield AL10 9AB, UK}
\ead{n.nishimura@keele.ac.uk (expired); nobuya.nishimura@yukawa.kyoto-u.ac.jp}

\newcommand{\aap}{A\&A}
\newcommand{\apj}{ApJ}
\newcommand{\apjs}{ApJS}
\newcommand{\apjl}{ApJL}
\newcommand{\mnras}{MNRAS}

\begin{abstract}
%% pourpose
The s-process in massive stars, producing nuclei up to $A\approx 90$,
has a different behaviour at low metallicity if stellar rotation is significant.
%The progress of nucleosynthesis is enhanced,
%and can produce heavier s-process elements $A \sim 90$--$140$.
This {\it enhanced s-process} is distinct from the s-process in massive stars 
around solar metallicity,
and details of the nucleosynthesis are poorly known.
%% method
We investigated nuclear physics uncertainties in the enhanced s-process in metal-poor stars within a Monte-Carlo framework.
We applied temperature-dependent uncertainties of reaction rates, distinguishing contributions from the ground state and from excited states.
%% results
We found that the final abundance of several isotopes shows uncertainties larger than a factor of 2,
mostly due to the neutron capture uncertainties.
A few nuclei around branching points are affected by uncertainties in the $\beta$-decay.
%% conclusion
%\textcolor{red}{The further studies are needed for the understanding of enhanced s-process in low metal poor stars.}
\end{abstract}

%%%%%%%%%%%%%%%%%%%%%%%%%%%%%%%%%%%%%%%%%
%     SECTION  1                                                                                                                    %
%%%%%%%%%%%%%%%%%%%%%%%%%%%%%%%%%%%%%%%%%

\section{Enhancement of weak s-process by stellar rotation}
%RH > --> \gtrsim
The s-process in massive stars ($\gtrsim 10M_\odot$)
is called ``weak s-process'',
because the major products are limited to lighter s-process elements
up to $A \approx 90$ (heavier elements up to Pb and Bi are produced in low mass AGB stars,
called {\it main s-process}, see \cite{2011RvMP...83..157K} for a review).
It takes place during the He-core and C-shell burning,
and the main neutron source is an $\alpha$-capture reaction ${}^{22}$Ne($\alpha$, n)${}^{25}$Mg
following the reaction sequence
${}^{14}{\rm N}(\alpha,\gamma){}^{18}{\rm F}(\beta^-){}^{18}{\rm O}(\alpha,\gamma){}^{22}{\rm Ne}$.
For very metal-poor stars, the behaviour of this weak s-process drastically changes due to rotation, and heavy nuclei with $A \approx 138$, including Ba, are produced
\cite{2008ApJ...687L..95P, 2012A&A...538L...2F, 2016MNRAS.456.1803F}.
Rotation-induced 
%RH strong 
mixing between the helium-burning and hydrogen-burning convective zones increases the abundance of primary $^{14}$N (and thus of ${}^{22}$Ne).
This increases the neutron captures and enhances the weak s-process
(hereafter, we call it {\it enhanced s-process}, ``e.s-process'').

\begin{figure}
\begin{center}
\includegraphics[width=\hsize]{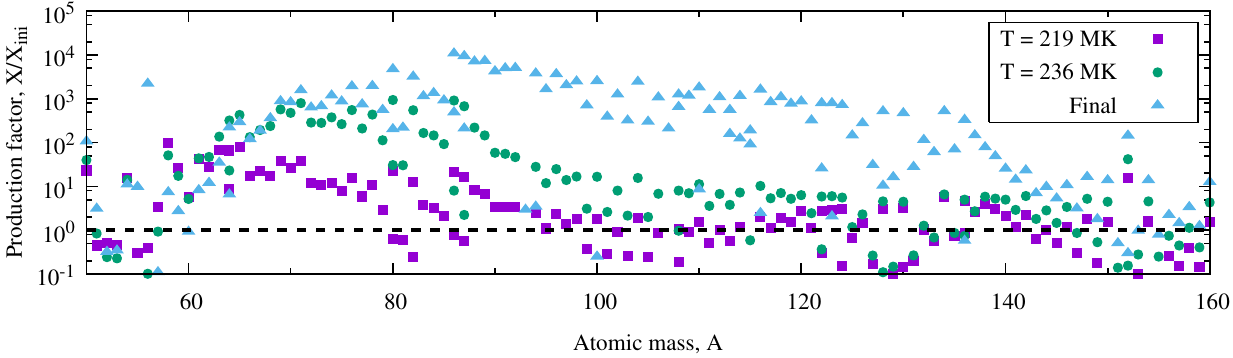}
\end{center}
\caption{Evolution of abundances in the e.s-process with different temperatures.}
\label{fig-sproc}
\end{figure}

\begin{figure}[b]
\begin{center}
\includegraphics[width=0.49\hsize]{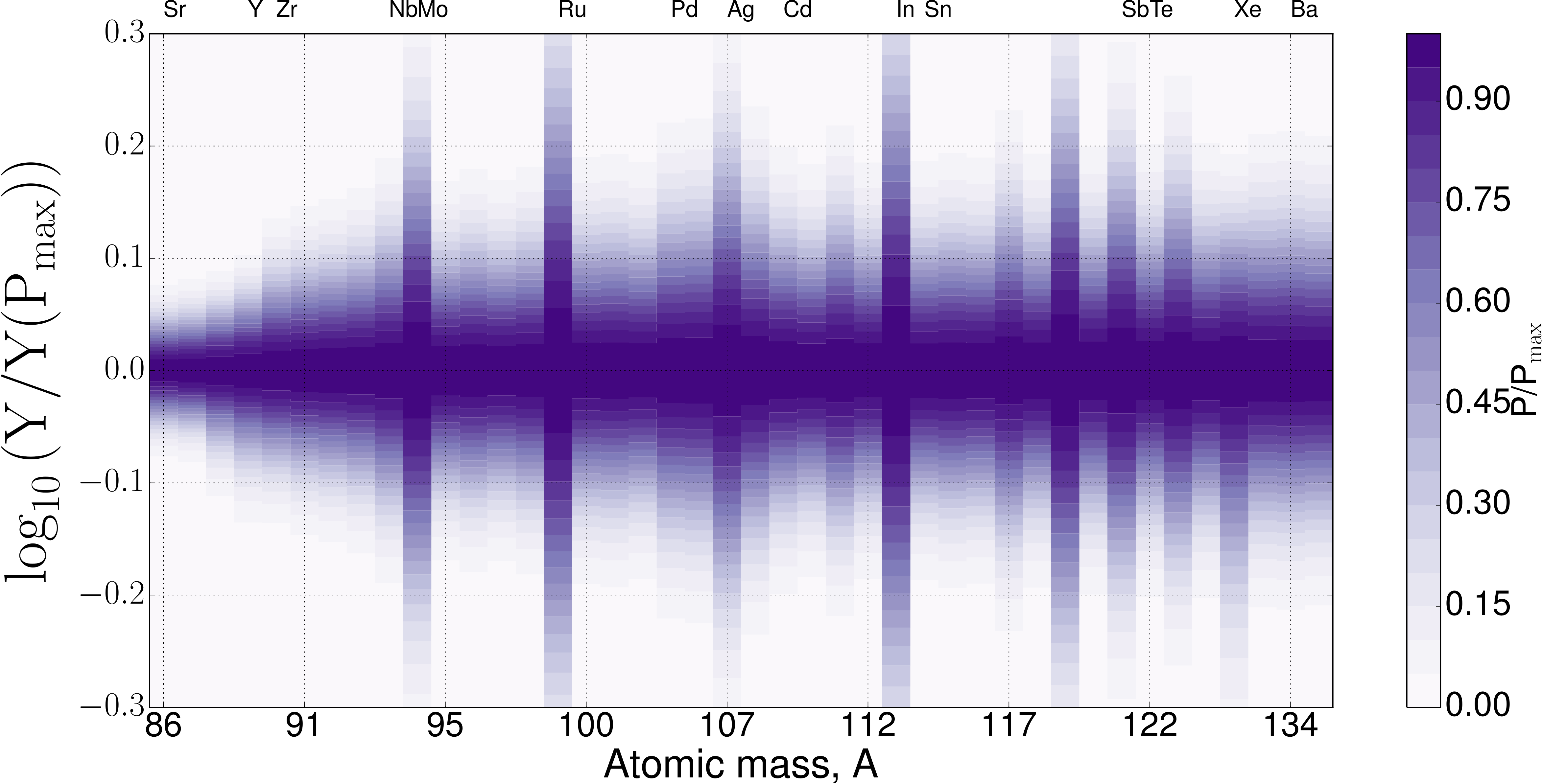}
\includegraphics[width=0.49\hsize]{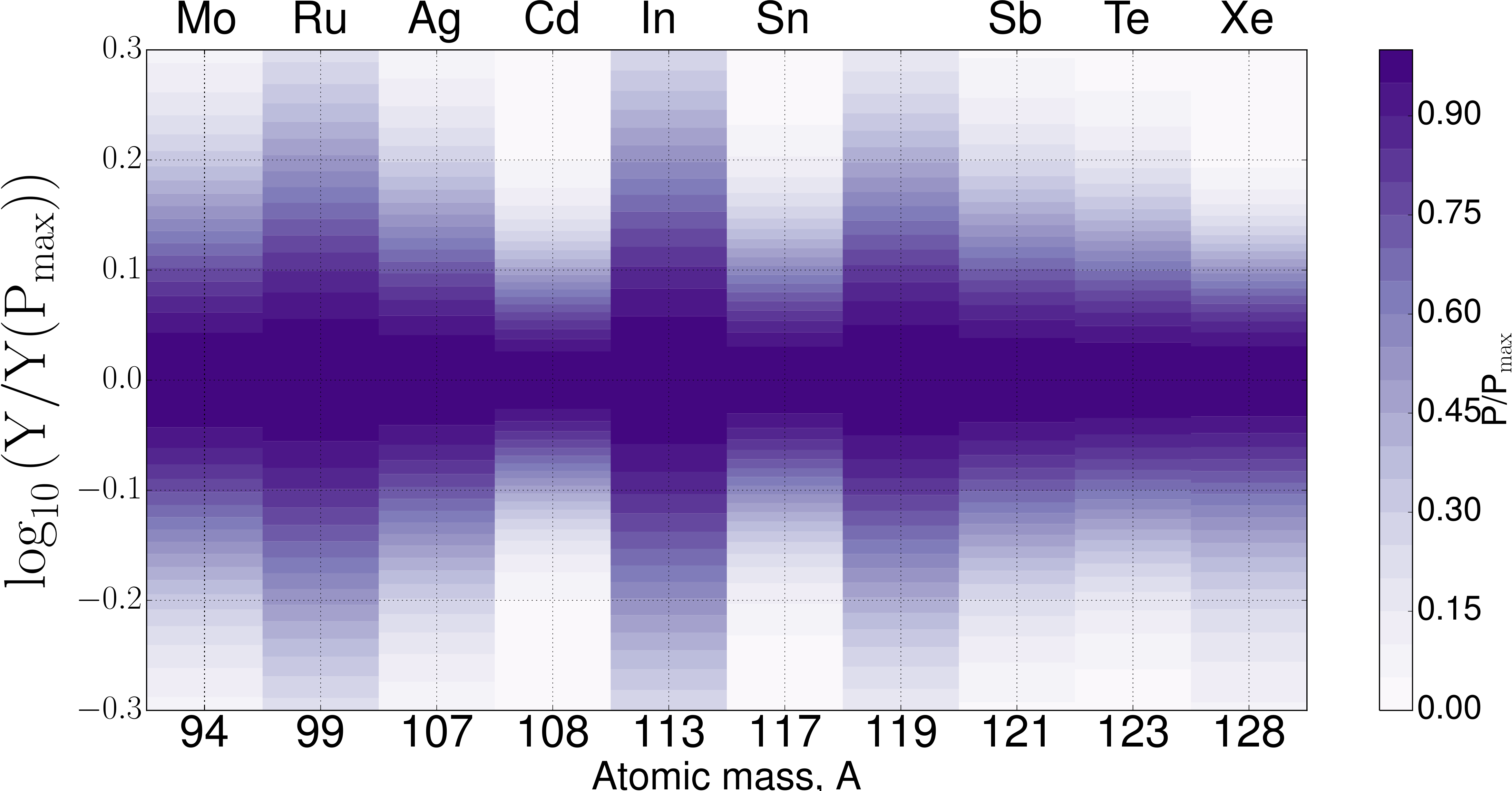}
\end{center}
\caption{
Uncertainties in final isotope production when all reactions are varied
for stable isotopes that e.s-process produces (left) and selected isotopes with larger uncertainty (right).
For each isotope, the normalized probability density distribution $Y/Y(P_{\rm max})$ is shown.}
\label{fig-mc-all}
\end{figure}

%A abundance evolution history of the e.s-process is shown in the left of 
The evolution of the e.s-process pattern is shown in Figure~\ref{fig-sproc}.
This is based on a simplified evolutionary track from a $25 M_\odot$ model,
effectively accounting for rotation-induced mixing by increasing the primary $^{14}$N,
as introduced in a previous study \cite{2014AIPC.1594..146N}.
The abundance pattern at $T=219$~MK is similar to the weak s-process
producing isotopes up to $A=90$.
As temperature increases ($T=236$~MK),
the distribution goes to the higher $A$ region
and finally reaches $A \approx 140$.
The final abundances show production in the range $90\leq A \leq 140$.

%The nucleosynthesis flow of the e.s-process
%is shown in the right of Figure~\ref{fig-sproc}.
%The main path of the e.s-process follows the line of stability, as also seen in the regular s-process.
%The flow reaches isotopes with $A>100$,
%which is not found in the regular weak s-process.
%The final abundances in the e.s-process are affected by (n,$\gamma$) and $\beta$-decay
%of isotopes at $A>100$.

%%%%%%%%%%%%%%%%%%%%%%%%%%%%%%%%%%%%%%%%%
%     SECTION  2                                                                                                                    %
%%%%%%%%%%%%%%%%%%%%%%%%%%%%%%%%%%%%%%%%%

\section{Uncertainty of reaction rates and Monte-Carlo simulation}

Nucleosynthesis in the e.s-process is quantitatively different from the standard weak s-process.
The impact of reactions relevant to stellar burning ($\alpha$-captures) of lighter nuclei
has been studied before, but with a focus on neutron-source and -poison reactions \cite{2014AIPC.1594..146N}.
However, the impacts of uncertainties in (n,$\gamma$) and $\beta$-decay
on the path of the e.s-process are poorly known.
We investigated, therefore, the role of nuclear physics input on the path of the e.s-process.
For this purpose, we performed Monte-Carlo (MC) simulations focusing on (n,$\gamma$) and $\beta$-decay reactions.
We use a MC framework with a general reaction network which is applicable to a variety of nucleosynthesis processes \cite{2014nic..confE.127N, 2014nic..confE.141R}.
%The Monte-Carlo code contains general nuclear reaction network, which applicable any kind of nucleosynthesis.

The input of nuclear physics uncertainty (i.e., of reaction rates)
is crucial for studying impacts on the nucleosynthesis yields.
In the present study, we assumed that reaction rates
have a temperature-dependent uncertainty because the relative contributions by the ground state (g.s.) and excited states to the rate change with
temperature and experimental cross sections, if available at all, only constrain g.s.\ contributions.
Following the prescription in \cite{2011ApJ...738..143R, 2012ApJS..201...26R},
experimental uncertainties are used for the g.s.\ contributions to (n,$\gamma$) rates,
whereas a factor $2$ is used for excited state uncertainties.
We simply apply a constant value $2$ for theoretical rates.
A similar approach is used for $\beta$-decay rates,  based on partition functions to determine the importance of excited states.
The uncertainty at lower temperatures ($T < 10^7$~K) corresponds to the one of measured decays,
while the uncertainty becomes larger as the temperature increases (for details, see \cite{2012ApJS..201...26R}).
A uniform random distribution between the upper and lower limit of the reaction rate at a given temperature was used for the MC variation factors.

\begin{figure}[t]
\begin{center}
\includegraphics[width=0.49\hsize]{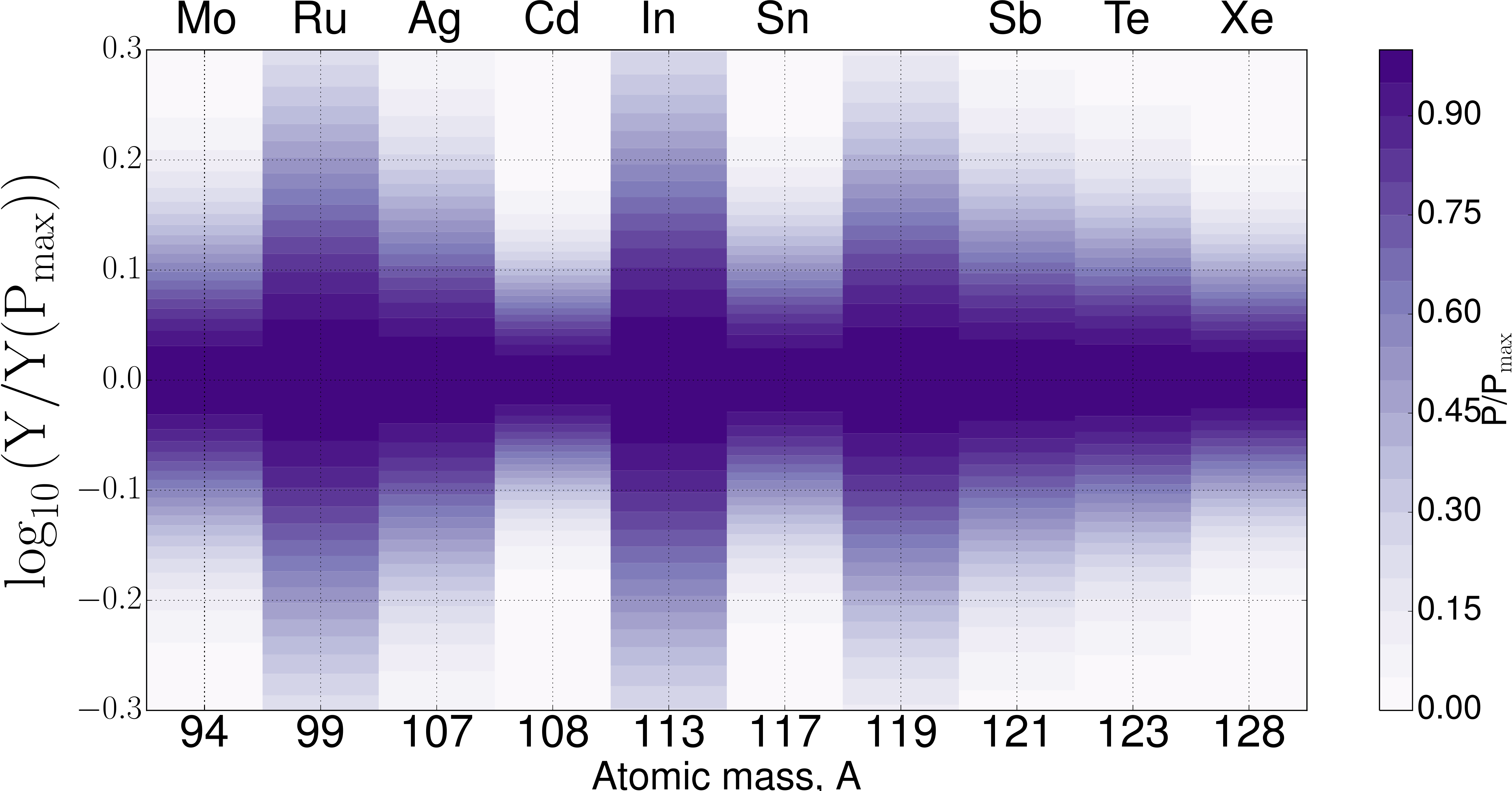}
\includegraphics[width=0.49\hsize]{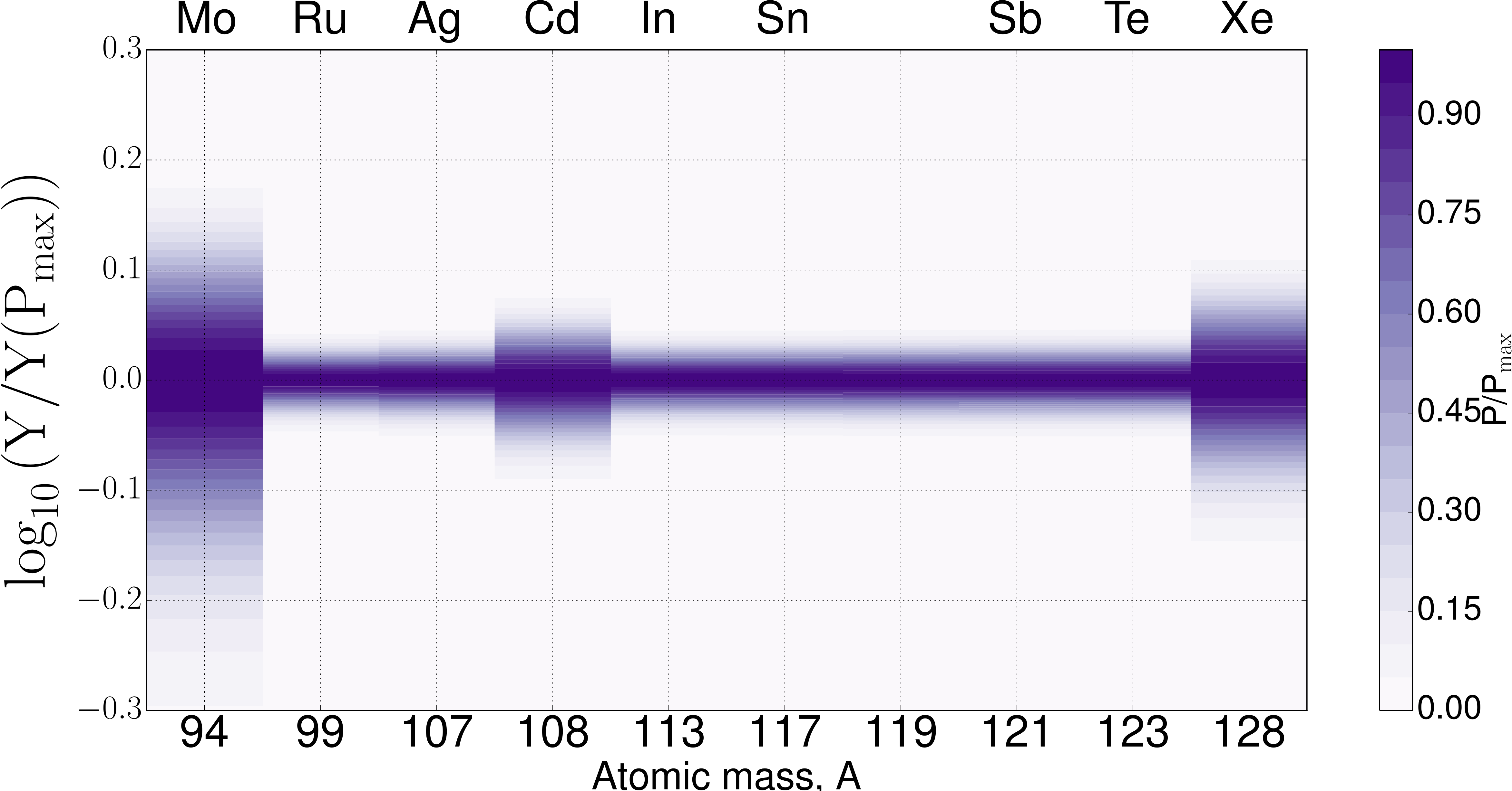}
\end{center}
\caption{Same as Figure~\ref{fig-mc-all}, where only (n,$\gamma$) (left) and only $\beta$-decay (right) are varied.}
\label{fig-mc-sep}
\end{figure}

Figure~\ref{fig-mc-all} shows the resulting production uncertainty
for the cases where we varied all (n,$\gamma$) reactions and $\beta$-decays.
We chose to show abundance uncertainties for stable s-process isotopes with $86\leq A\leq 136$ in the left panel,
which cover the main products of the e.s-process.
We plot isotopes showing the largest uncertainties in the right panel.
The colour distribution corresponds to the normalized probability density distribution of the uncertainty in the final abundance.
Values of $0.114$ and $0.301$ for the probability density distribution correspond to $30$~\% and a factor of $2$ uncertainty, respectively.

Some isotopes show larger uncertainties, while most others are within the $\pm 30$~\% range.
In order to identify the role of the reactions for the final uncertainty, we varied neutron captures and $\beta$-decays separately.
Figure~\ref{fig-mc-sep} shows the results for the isotopes with the largest uncertainties,
the left and right panels correspond to variation of all (n,$\gamma $) and all $\beta$-decays, respectively.

The results clearly indicate that (n,$\gamma$) reactions have a dominant impact on the nucleosynthesis uncertainty
and that $\beta$-decays have a limited importance:
(i) the uncertainties in (n,$\gamma$) leads to a general $30$~\% uncertainty in the final abundances, several isotopes having a higher value up to a factor of 2;
(ii) $\beta$-decays only affect a few isotopes around the branching points at $A \approx 94$, $108$, and $128$.
%RH: I do not see what this sentence adds?
%These MC results clearly show individual impacts of (n,$\gamma$) and $\beta$-decay on the final uncertaitny.

%%%%%%%%%%%%%%%%%%%%%%%%%%%%%%%%%%%%%%%%%
%     SECTION  3                                                                                                                    %
%%%%%%%%%%%%%%%%%%%%%%%%%%%%%%%%%%%%%%%%%

\section{Conclusion}

In this study, we evaluated the impact on e.s-process nucleosynthesis of nuclear physics uncertainties
using MC calculations.
The method can identify the importance of reactions
and we found that (n,$\gamma$) reactions dominate the total uncertainty,
with a few important contributions from $\beta$-decays around branching points.
Our method is a robust way to identify key reaction rates to support further investigations in nuclear astrophysics
regarding the e.s-process.

We were supported by the BRIDGCE UK (www.astro.keele.ac.uk/bridgce),
the ERC (EU-FP7-ERC-2012-St Grant 306901, EU-FP7 Adv Grant GA321263-FISH),
and the UK STFC (ST/M000958/1), COSMOS (STFC DiRAC Facility)
at DAMTP in University of Cambridge.

\section*{References}

\bibliographystyle{iopart-num.bst}
\bibliography{ref.bib}

%\begin{thebibliography}{9}
%\bibitem{iopartnum} IOP Publishing is to grateful Mark A Caprio, Center for Theoretical Physics, Yale University, for permission to include the {\tt iopart-num} \BibTeX package (version 2.0, December 21, 2006) with  this documentation. Updates and new releases of {\tt iopart-num} can be found on \verb"www.ctan.org" (CTAN). 
%\end{thebibliography}

\end{document}